\documentstyle[epsfig]{cupconf}

\ifoldfss
\else
  \ifnfssone
    \newmathalphabet{\mathit}
      \addtoversion{normal}{\mathit}{cmr}{m}{it}
      \addtoversion{bold}{\mathit}{cmr}{bx}{it}
    \newmathalphabet{\mathcal}
      \addtoversion{normal}{\mathcal}{cmsy}{m}{n}
    \else
    \ifnfsstwo
    \fi
  \fi
\fi

\def\nuc#1#2{\relax\ifmmode{}^{#1}{\protect\text{#2}}\else${}^{#1}$#2\fi}
\def\ltsima{$\; \buildrel < \over \sim \;$}
\def\simlt{\lower.5ex\hbox{\ltsima}} 
\def\gtsima{$\; \buildrel > \over \sim \;$}
\def\simgt{\lower.5ex\hbox{\gtsima}} 

\def\hexnumber#1{\ifcase#1 0\or1\or2\or3\or4\or5\or6\or7\or8\or9\or
 A\or B\or C\or D\or E\or F\fi }
\makeatletter
\ifx\CUP@mtlplain@loaded\undefined
\else
\fi
\makeatother
 \makeatletter
 \ifx\CUP@mtlplain@loaded\undefined
   \font\tenbmi=cmmib10 at 10pt
   \font\sevenbmi=cmmib10 at 7pt
   \font\fivebmi=cmmib10 at 5pt

   \newfam\bmifam
   \textfont\bmifam=\tenbmi
   \scriptfont\bmifam=\sevenbmi
   \scriptscriptfont\bmifam=\fivebmi
   
 \fi
 \makeatother
\ifnfsstwo

\fi
\ifnfssone

\fi
\ifoldfss

\fi

\mathchardef\varLambda="0103

\makeatletter
\ifx\CUP@mtlplain@loaded\undefined
\else
\fi
\makeatother
\makeatletter
\ifx\CUP@mtlplain@loaded\undefined
  \font\tenbms=cmbsy10
  \font\sevenbms=cmbsy10 at 7pt
  \font\fivebms=cmbsy10 at 5pt
  \newfam\bmsfam
  \textfont\bmsfam=\tenbms
  \scriptfont\bmsfam=\sevenbms
  \scriptscriptfont\bmsfam=\fivebms

  \edef\bsy@{\hexnumber\bmsfam}
  \mathchardef\bnabla="0\bsy@72
\fi
\makeatother
\def\etal{\mbox{\it et al.}}

\title[The Biggest Explosion Since the Big Bang]{X-ray Emission of Supernova
1998bw in the Error Box of GRB980425}

\author[E. Pian]%
{E\ls L\ls E\ls N\ls A\ns P\ls I\ls A\ls N$^1$}

\affiliation{$^1$Istituto di Tecnologie e Studio delle
Radiazioni Extraterrestri, Via Gobetti 101, I-40129 Bologna, Italy}

\begin{document}
\ifnfssone
\else
  \ifnfsstwo
  \else
    \ifoldfss
      \let\mathcal\cal
      \let\mathrm\rm
      \let\mathsf\sf
    \fi
  \fi
\fi

\maketitle

\begin{abstract}

The spatial and temporal coincidence of a GRB and a supernova explosion
(1998bw) on 25 April 1998 has raised conjectures on the physical connection
between the two phenomena, and in general on the association of GRBs with
supernovae, at least with the most powerful among them (hypernovae or
collapsars). In fact, multiwavelength observations of SN~1998bw have revealed
unusual characteristics: extremely high energy output at radio and optical
wavelengths, and relativistic expansion of the outgoing shock.  The X-ray
emission of SN~1998bw, monitored by BeppoSAX starting $\sim$10 hours after the
GRB detection, was remarkably prompt (within one day of supernova detonation),
but exhibited spectral and temporal properties similar to those of other
supernovae detected in X-rays.

\end{abstract}

\firstsection 

\section{Introduction}

The discovery of a supernova within the 8$^{\prime}$ radius error circle of the
GRB980425 has been regarded as a major puzzle within the thick mystery of GRBs. 
The GRB980425, which has been detected by the BeppoSAX Gamma Ray Burst Monitor
(GRBM, 40-700 keV) and by BATSE (Kippen 1998), and rapidly localized by the
BeppoSAX Wide Field Cameras (WFC, 2-26 keV) Unit 2, appears as a relatively
weak burst, characterized by a single, non structured peak of longer duration
in the 2-26 keV range (52 seconds) than in the 40-700 keV range (31 seconds). 
In Figure 1 (left panel)  are reported the temporal profiles in both energy
ranges.  The spectrum of the GRB rapidly softens with time (Figure 1, right
panel; see also Frontera et al. 1999). 

Weak intensity, a single peak, a soft and fastly evolving spectrum, and a
$\sim$5 second temporal delay of the X-rays with respect to the
$\gamma$-rays appear to be the main characteristics of this GRB.  However,
these features are common to other GRBs, and therefore cannot be
considered as an obvious suggestion that GRB980425 is peculiar.

It came as a surprise for the teams involved in GRB search and follow-up at
longer wavelengths, and subsequently for the whole astronomical community,
that in the error box of GRB980425 a supernova was detected, 1998bw, at the
optical (Galama et al.  1998; Galama et al. 1999a), and radio wavelengths
(Kulkarni et al.  1998a), 17 hours and 3 days after the GRB event,
respectively. The inferred time of supernova explosion is consistent with the
GRB occurrence to within +0.7/-2 days (Iwamoto et al. 1998).

SN~1998bw lies in a spiral arm of the galaxy ESO 184-G82, at a redshift $z =
0.0085$ (Tinney et al.  1998). Its radio luminosity ($\sim
10^{38}$ erg s$^{-1}$ at peak) is the largest ever measured for any supernova,
and the optical one ($\sim 10^{43}$ erg s$^{-1}$ at peak) ranks among the
highest supernova luminosities.

SN~1998bw stands out not only for its positional and temporal coincidence
with GRB980425 and for its unusual radio and optical luminosity, but also for
the properties of the radio light curves (Kulkarni et al. 1998a; Wieringa et
al.  1999) and for the broad optical spectral lines, which indicate high
photospheric velocities. Based on its optical spectrum, SN~1998bw was
classified as a peculiar Type Ib before maximum light (Sadler et al. 1998) 
and Type Ic at later epochs (Iwamoto et al. 1998; Galama et al. 1998;  Patat
\& Piemonte 1998).

\section{BeppoSAX Target of Opportunity Observations of the GRB980425 Error
Box}

Following the GRB event, the field of GRB980425 was promptly acquired by the
BeppoSAX Narrow Field Instruments (NFI; these include the LECS, MECS, HPGSPC,
and PDS detectors. See Butler \& Scarsi 1990, and Boella et
al.  1997 for a description of the BeppoSAX mission), and observations in the
energy range 0.1-300 keV started 10 hours after the GRB detection (26-27
April 1998). Two previously unknown sources have been detected within the WFC
error box by the MECS in the 2-10 keV energy range (Pian et al. 
1999a;  Pian et al. 1999b).

The brighter source, S1, is consistent with the position of SN~1998bw, while
the fainter one, S2, is not (Figure 2).  The LECS data have a significantly
lower
signal-to-noise ratio than the MECS data and the HPGSPC and PDS instruments
yielded no detection above the background, therefore we will briefly report
here only the MECS results and refer to a paper of imminent submission for
details about both LECS and MECS data (Pian et al 1999b).

Note that the coordinates of sources S1 and S2 distributed by Pian et al.
(1998)  have been revised in November 1998, to take into account a systematic
error due to the non-optimal spacecraft attitude during the April and May 1998
observations (see Piro et al. 1998a).  Figure 2 illustrates
the updated, correct location of the sources. 

The following NFI pointings, one week (2-3 May 1998) and six months
later
(10-12 November 1998), have shown that neither source exhibits the behavior
expected for an X-ray afterglow. Source S1 did not exhibit significant
variability in one week, and was still detected, a factor of $\sim$2 fainter,
six months later (see Figure 3).  Source S2 exhibits marginally significant
variability between 26-28 April and 2-3 May 1998.  It is not detected in
November 1998, but its upper limit is consistent with the April-May flux level
(see Figures 3 and 4a). 

The variability of source S1 and its positional consistency with SN~1998bw
suggest that S1 is the X-ray counterpart of the supernova.  This is the
earliest detection of X-ray supernova emission, and the first detection of
medium energy X-rays from a Type I supernova (the only other case of X-ray
bright supernova is the Type Ic SN~1994I, detected in the soft X-rays by
ROSAT, Immler et al. 1998a). 

At the distance of SN~1998bw, the luminosity observed in the range 2-10
keV, $\sim$2-5$\times 10^{40}$ erg s$^{-1}$, is compatible with that of
other supernovae detected in the same energy band, all Type II (see
Table~1).  However, the supernova X-ray luminosity could suffer from host
galaxy contamination, which might be significant at these energies (see
Fabbiano 1989).  Similarly, the observed variation of a factor of two in
six months is only a lower limit to the supernova X-ray variability
amplitude.  A power-law $f(t) \propto t^{-p}$ with index $p \sim
0.2$ provides an acceptable fit of the light curve (Fig. 4b), and is
approximately consistent with the behavior observed for other supernovae
(Kohmura et al.  1994; Houck et al. 1998)  and
with predictions based on interaction of energetic electrons with the
circumstellar medium (Chevalier \& Fransson 1994; Li \& Chevalier 1999).

The prompt X-ray emission observed for SN~1998bw requires that the
circumstellar medium is highly ionized (probably by the powerful   
explosion), to allow the X-rays to escape so soon after the explosion (see
Zimmermann et al. 1994), and also very dense, as inferred also from the
large radio output (Kulkarni et al. 1998a;  Wieringa et al. 1999).

The spectrum of S1 in the 2-10 keV energy range is well fitted by a
power-law $F_\nu \propto \nu^{-\alpha}$ of index $\alpha = 0.5 \pm 0.2$
(1-$\sigma$), or by a thermal bremsstrahlung model with temperature
$\sim$15 keV (see Pian et al. 1999b for details on the spectral fits).
Both are consistent with spectral slopes and temperatures found for other
supernovae detected in X-rays (e.g., Kohmura et al. 1994; Leising et al.
1994; Dotani et al.  1987). 

The mildly relativistic conditions of the expanding shock of SN~1998bw
(Kulkarni et al. 1998a) might suggest that the mechanism responsible for
the X-ray emission is synchrotron radiation of very energetic electrons,
or inverse Compton scattering of relativistic electrons (which produce the
radio spectrum via synchrotron) off optical/UV photons of the thermal
ejecta. The X-ray spectral index is consistent with that measured for the
radio spectrum starting $\sim$15 days after the explosion (Kulkarni et al.
1998a;  Wieringa et al. 1999;  before that epoch the radio spectrum is
significantly self-absorbed), and with the spectral slope connecting
quasi-simultaneous radio and X-ray measurements ($\alpha \sim 0.8$). 
Therefore, in case the X-rays have a non-thermal origin, it is difficult
to establish whether they are produced through the synchrotron or inverse
Compton process (Fig. 5). 

\section{The GRB/Supernova Connection}

Although the chance coincidence of GRB980425 and SN~1998bw has a very low
probability ($10^{-4}$, Galama et al. 1998), the GRB community has not
accepted unanimously the physical association of the GRB and the
supernova. In fact, the faint source S2 - possibly, but not clearly,
fading - could be considered an afterglow candidate. The flux of S2 during
the first BeppoSAX observation would be consistent with a power-law decay
of index $p \simeq 1.3$ after the early X-ray emission observed by the
WFC
(see Fig. 4a).  This is in the range of the power-law decay indices of
``classical" X-ray afterglows (Costa et al.  1997;  Nicastro et al. 1998; 
Dal Fiume et al. 1999; in 't Zand et al. 1998; Nicastro et al.  1999; 
Vreeswijk et al. 1999a;  Heise et al. 1999).  However, the second
detection of S2 is not conclusive: it is marginally consistent with the
first detection, but it is also marginally consistent with the power-law
decay. The November 1998 upper limit is consistent with the detection
level. Therefore, based on the present data, one cannot establish whether
S2 is an afterglow exhibiting a small re-bursting (similar to GRB970508,
although the time scale would be different, Piro et al. 1998b) or a
permanent, perhaps modestly variable, X-ray emitter, like an active
galactic nucleus or a Galactic binary (the chance probability of detecting
a source of the level of S2 in the $8^{\prime}$ radius WFC error box of
GRB980425 is rather high, $\sim$12\%).  Optical observations have been
equally inconclusive: no optical transient at a position consistent with
S2 has been detected by early imaging of the GRB error box, and late epoch
optical spectroscopy of sources brighter than $V \sim$18 in the S2 error
box failed to identify any active galaxy or binary stellar system having a
compact object (Halpern 1998, and private communication). 

At the time of GRB980425/SN~1998bw detection, five optical afterglows of GRBs
had been detected, and for all of them, similarly to X-ray afterglows, a
rapid power-law decay had been measured with index $p$ in the range 1.1-2.1
(Van Paradijs et al. 1997;  Fruchter et al. 1999a; Fruchter et al. 1999b, and
references therein; Diercks et al. 1998; Halpern et al. 1998; Kulkarni et al. 
1998b; Groot et al. 1998; Palazzi et al.  1998).  The circumstance of
detecting a supernova as the possible counterpart of a GRB was unprecedented. 
Therefore, it was proposed that this GRB might belong to a different class of 
events, with apparently indistinguishable high energy characteristics, but
with different progenitors. Furthermore, assuming association with
SN~1998bw,
GRB980425 would be much closer than the GRBs for which a redshift measurement
is available, which reinforced the idea that GRB980425 was physically
dissimilar from GRBs exhibiting power-law decaying X-ray
and optical remnants, predicted by the cosmological fireball model (Rees \&
M\'esz\'aros 1992; Piran 1999). 

After the case of GRB980425/SN~1998bw, many authors have searched
for
statistical support of the possible association between GRBs and supernovae,
and obtained different, and sometimes conflicting, results.  The comparison of
the BATSE catalog with supernovae compilations seems to suggest that some GRBs
may be spatially (within an angular uncertainty of many degrees) and
temporally (within $\sim$ 20-30 days) consistent with Type Ib/c supernovae,
while association with Type Ia is ruled out (Wang \& Wheeler 1998. See however
Kippen et al. 1998 and Graziani et al. 1998).  Association has been
specifically
proposed for the cases of the Type II supernovae 1997cy and 1999E with
GRB970514 and GRB980910, respectively, based on temporal and spatial proximity
and on the outstanding optical properties of the two supernovae (Woosley et
al. 1999; Germany et al. 1999; Thorsett \& Hogg 1999;  Turatto et al. 1999). 
However, limiting the GRB sample to the events with temporal profile similar
to GRB980425 leads to no significant association (Bloom et al.  1998a). A
negative result is also obtained by further restricting the subset to long,
soft GRBs (Norris et al. 1998). This seems to suggest that the temporal and
spectral characteristics of GRBs are not obvious tracers of possible
association with supernovae.

More recent studies have shown that the optical afterglows of some GRBs
exhibit deviations from a ``pure" power-law decay, and these have been
ascribed to the possible presence of a supernova underlying the afterglow
(GRB970228, Reichart 1999; Galama et al. 1999b; GRB970508, Germany et al. 
1999; GRB980326, Bloom et al. 1999; GRB990510, Fruchter et al. 1999c;
Beuermann et al. 1999; GRB990712, Hjorth et al. 1999). This makes the
association between GRB980425 and SN~1998bw more solid, and supports the
speculation that all GRBs of long duration ($>$1 s) are formed by
extremely energetic supernova explosions (``failed" supernovae,
hypernovae, or collapsars, Paczy\'nski 1998; Woosley et al. 1999;
MacFadyen \& Woosley 1998). These observational hints and theoretical
picture suggest that GRB980425 and some GRBs for which a counterpart has
been detected at frequencies lower than the $\gamma$-rays belong to a
same class and have similar progenitors, despite the different distance
and behavior of the multiwavelength counterparts. 

Indeed, the recent discovery of a GRB optical afterglow at the
intermediate redshift $z = 0.43$ (GRB990712, Galama et al. 1999c) might
support a continuity of properties between GRB980425 and the other
precisely localized GRBs, perhaps based on the different amount of
beaming, according to the degree of jet alignment (Eichler \& Levinson
1999; Cen 1998; Postnov et al. 1999;  Woosley et al. 1999). In this
scenario, in highly beamed GRBs the non-thermal multiwavelength afterglow
could overwhelm the underlying supernova emission.  The latter should
instead be detected more clearly in GRBs seen off-axis, like GRB980425,
which also appear weaker.  Assuming association with SN~1998bw and
isotropic emission, the total energy of GRB980425 in the 40-700 keV range,
$\sim 5 \times 10^{47}$ erg, is at least four order of magnitudes less
than that of GRBs with known distance (see Figure 6).

\section{X-ray Supernovae and Gamma-Ray Bursts}

GRB980425 has become a milestone in the history of GRB research in that it
provided a strong suggestion toward the determination of GRB
progenitors.
The BeppoSAX
rapid turnaround allowed the most prompt detection ever of X-rays from a
supernova, thus bringing to 10 the number of supernovae detected at these
energies (barring supernova remnants).  The complete list is reported in
Table 1, which represents an
update of Table 3 in the review by Schlegel (1995).  In addition, X-ray
luminosities at the discovery epoch are reported. 

Since the epoch of Schlegel's review, the number of X-ray supernovae has
doubled, and the detection of SN~1998bw has confirmed that medium energy
X-rays are produced also from Type Ib/c supernovae.  This result was
predictable, given that these must have environments similar to Type II,
namely a dense circumstellar medium produced by the slow wind of the
progenitor, with which the supernova shock interacts producing both radio and
X-ray emission. 

Several issues have still to be clarified about GRB980425/SN~1998bw for what
concerns X-ray emission.  Particularly, an observation of the field with an
instrument with good imaging angular resolution (like Chandra) is required to
study in detail the host galaxy in X-rays and to disentangle it from the
point-like supernova emission at the various epochs of BeppoSAX observation.
A very sensitive instrument (like XMM) would instead allow a deep survey of
the field of GRB980425 in medium energy X-rays, to detect the weak source S2
with a good signal-to-noise ratio, assuming it is broadly constant in the
long term.  (Its non detection would be perhaps more constraining toward its
identification with the GRB afterglow.) 

Based on the recent findings of possible supernova emission underlying the
optical, power-law fading remnants of some GRBs, future research should
exploit the X-ray observing facilities in search of analogous signatures
in the X-ray afterglows.  If GRBs are produced by supernovae, the same
conditions which make detectable the afterglow, i.e. the presence of a
sufficiently dense medium, should also favor the production of X-rays from
the supernova.  A possible past example might be the re-bursting of
GRB970508 (Piro et al.  1998b), but it would imply a supernova X-ray
luminosity four orders of magnitude larger than that of the most luminous
X-ray
supernova so far detected, 1988Z (Table 1). Therefore, the X-ray data
available to date do not suggest any evidence of a supernova underlying a
GRB afterglow.  Clearly, more GRB localizations and observations of
targets at redshifts no larger than $\sim$0.1 are necessary to make a
significant supernova detection affordable by the presently available
X-ray instruments.

\begin{acknowledgments}

I am grateful to L.  Amati, A. Antonelli, F. Boffi, R. Chevalier, E. Costa, J. 
Danziger, F.  Frontera, A. Fruchter, T. Galama, P. Giommi, J. Halpern, J. 
Katz, S.  Kulkarni, N. Masetti, E. Palazzi, N.  Panagia, S. Perlmutter, 
L.  Stanghellini, M.  Turatto, P. Vreeswijk, C.  Wheeler, T. Young for
valuable comments and many technical and scientific inputs.  I would like
to thank Mario Livio and the other STScI May Symposium organizers for a
pleasant and stimulating conference. 

\end{acknowledgments}

\newpage
\null
\vspace{1cm}

\centerline{\underline{\bf Figure Captions}}

\vspace{1cm}

{\bf Fig. 1:}
Left panel: BeppoSAX WFC (top) and GRBM (bottom) light curves
of GRB980425.  Time is in seconds; the onset of the GRB, indicated by the
zero abscissa, corresponds to 1998 April 25.9091.  The typical 1-$\sigma$
uncertainty associated with the individual flux points is $\sim$4 counts
s$^{-1}$ for the WFC data and $\sim$40 counts s$^{-1}$ for the GRBM data.
The vertical dashed lines divide the burst duration in two time intervals
denoted by ``A" (12 s) and ``B" (40 s), over which the signal has been
integrated to construct 2-700 keV spectra. These are reported in the right
panel, along with the Band law fitting curves (Band et al.  1993).

\bigskip

{\bf Fig. 2:}
Digitized Sky Survey image ($16^\prime \times 16^\prime$) of
the WFC error box of GRB980425 (large circle of radius $8^\prime$).  The
left boundary of the IPN annulus is indicated as well as the error boxes
of the two NFI X-ray sources (dashed circles of radius $1^{\prime}.5$) and
the position of the SN~1998bw. The two X-ray sources S1 (1SAXJ1935.0-5248,
at the revised coordinates (J2000) $\alpha$ = 19$^{\rm h}$ 35$^{\rm m}$   
05.9$^{\rm s}$, $\delta$ = $-52^{\circ} 50^{\prime} 03^{\prime\prime}$)   
and S2 (1SAXJ1935.3-5252, at the revised coordinates $\alpha$ = 19$^{\rm  
h}$ 35$^{\rm m}$ 22.9$^{\rm s}$, $\delta$ = $-52^{\circ} 53^{\prime}
49^{\prime\prime}$), and SN~1998bw are consistent with the WFC and IPN
locations. (From Galama et al.  1999a, reprinted with permission of {\it
Astronomy \& Astrophysics}.)

\bigskip

{\bf Fig. 3:}
BeppoSAX MECS images of GRB980425 in April 1998 (left) and November
1998 (right). The data have been smoothed with a Gaussian function of
$1^{\prime}.5$ FWHM. Source S1 is visible in the image center; source S2 is   
toward the South-East, at $\sim$4$^{\prime}$ away from S1.

\bigskip

{\bf Fig. 4:}
(a) BeppoSAX MECS light curves in the 2-10 keV band of the
X-ray sources S1 (open squares) and S2 (filled circles) detected in the
GRB980425 field. The WFC early measurement in the same band is also shown
(star).  The zero point for the abscissa is 1998 April 25.9091.
Uncertainties associated with the WFC point and with the NFI measurements
of S1, being equal to or smaller than the symbol size, have been omitted.
The dotted line represents the power-law $f(t) \propto t^{-p}$ of index  
$p \simeq 1.3$ connecting the WFC measurement with the first NFI
measurement of source S2. The extrapolation of the line to the time of
the third observation falls below the lower bound of the S2 flux
measurement but it is marginally consistent with it (the excess with
respect to the power-law is $\sim$2.5-$\sigma$).  (b) Same as (a) for
source S1 only.  The fit to the temporal decay with a power-law of index
$\sim 0.2$ is shown as a dotted line.

\bigskip

{\bf Fig. 5:}
Quasi-simultaneous radio-to-X-ray spectral energy distributions of
SN~1998bw in 3-5 May (open circles) and 10-12 November 1998 (filled circles).
Power-law fits to the X-ray spectra are shown along with their 1-$\sigma$
confidence ranges (Pian et al. 1999b).
The optical magnitudes have been transformed to fluxes according to Fukugita
et al. (1995) and corrected for Galactic absorption using $A_V = 0.2$    
(Schlegel et al. 1998), although Patat et al.  (1999)  argue in favor of a
lower value.  For the first epoch, the optical and radio data have been taken
from Galama et al. (1998)  and Kulkarni et al. (1998a), respectively.  For the
second epoch the optical data have been either interpolated (bands $V$ and
$I$) between October 29 (McKenzie \& Schaefer 1999) and November 26 (Vreeswijk
et al. 1999c)  measurements or extrapolated (bands $B$ and $R$) using the late
time exponential decay fitted to the light curves by Patat et al. (1999). The
radio data are from Wieringa et al. (1999). Note that the BeppoSAX data 
represent the blend of the supernova and possible host galaxy emission and
should then be considered upper limits on the X-ray emission of SN~1998bw.
The optical supernova ejecta dominate the power output at both epochs.  The
radio and X-ray data could be consistent with a single radiation component.

\bigskip

{\bf Fig. 6:}
GRB energy output in the 40-700 keV range (assumed to be emitted
isotropically) vs redshift.  The open circle corresponds to GRB980329, for
which the redshift was only estimated with arguments based on the appearance
of the optical spectrum (Fruchter 1999.  This value is controversial:     
the Keck detection of the possible host galaxy of the GRB would point to a
lower redshift, Fruchter 1999, private communication. See alternatively
Draine 1999). Redshift measurements are
from Djorgovski et al. (1999a); Bloom et al. (1998b);  Kulkarni et al.
(1998b); Tinney et al. (1998); Djorgovski et al. (1999b); Djorgovski et al.
(1998); Kulkarni et al. (1999); Vreeswijk et al. (1999b).  The references for
$\gamma$-ray fluences measured by the BeppoSAX GRBM are Frontera et al.    
(1998);  Piro et al. (1998b); Dal Fiume et al. (1999);  in 't Zand et al.
(1998);  Pian et al. (1999a);  Costa et al. (1999);  Amati et al. (1998); 
Feroci et al. (1999); Amati et al. (1999).
Adopted values for the Hubble constant and deceleration parameter are
$H_0 = 70$ km s$^{-1}$ Mpc$^{-1}$ and $q_0 = 0.15$, respectively.

\end{document}